\documentclass[12pt]{article}
\usepackage{amsmath} 
\input{epsf} 
\def\ltap{\raisebox{-.6ex}{\rlap{$\,\sim\,$}} \raisebox{.4ex}{$\,<\,$}}

\newcommand\as{\alpha_{\mathrm{S}}} 
 
\def\beq{\begin{equation}} 
\def\eeq{\end{equation}} 
\def\beeq{\begin{eqnarray}} 
\def\eeeq{\end{eqnarray}} 
 
\def\to{\rightarrow}

\def\WH{{\it WH}}
\def\ZH{{\it ZH}}
\def\VH{{\it VH}}

\begin{document} 

\begin{titlepage}
\begin{flushright}
CERN-PH-TH/2011-159\\
ZU-TH 14/11
\end{flushright}
\renewcommand{\thefootnote}{\fnsymbol{footnote}}
\par \vspace{10mm}

\begin{center}
{\Large \bf Associated \WH\ production at hadron colliders:}
\\[0.5cm]
{\Large \bf a fully exclusive QCD calculation at NNLO}
\end{center}
\par \vspace{2mm}
\begin{center}
{\bf Giancarlo Ferrera}$^{(a)}$, {\bf Massimiliano Grazzini}$^{(b)}$\footnote{On leave of absence from INFN, Sezione di Firenze, Sesto Fiorentino, Florence, Italy.}~~and~~{\bf Francesco Tramontano}$^{(c)}$\\

\vspace{5mm}

$^{(a)}$ Dipartimento di Fisica e Astronomia,
Universit\`a di Firenze and\\ INFN, Sezione di Firenze,
I-50019 Sesto Fiorentino, Florence, Italy

$^{(b)}$ Institut f\"ur Theoretische Physik, Universit\"at Z\"urich, CH-8057 Z\"urich, Switzerland

$^{(c)}$ Theory Group, Physics Department, CERN, CH-1211 Geneva 23, Switzerland

\vspace{5mm}

\end{center}

\par \vspace{2mm}
\begin{center} {\large \bf Abstract} \end{center}
\begin{quote}
\pretolerance 10000

We consider QCD radiative corrections to Standard Model Higgs boson production
in association with a $W$ boson in hadron collisions.
We present a fully exclusive calculation up to next-to-next-to-leading order (NNLO) in QCD perturbation theory. To perform this NNLO computation, we use a recently proposed version of the subtraction formalism. Our calculation includes finite-width effects, the leptonic decay of the $W$ boson with its spin correlations, and the decay of the Higgs boson into a $b{\bar b}$ pair.
We present selected numerical results at the Tevatron and the LHC.

\end{quote}

\vspace*{\fill}
\begin{flushleft}
July 2011

\end{flushleft}
\end{titlepage}

\setcounter{footnote}{1}
\renewcommand{\thefootnote}{\fnsymbol{footnote}}

The search for the Higgs boson
or its alternatives in theories beyond the Standard Model (SM)
is the key endeavour
of current high-energy colliders.
The Tevatron has recently excluded at the 95\% CL a SM Higgs boson in the
mass range $158$ GeV $<m_H<173$ GeV \cite{Aaltonen:2011gs}.
At the LHC the Higgs search has now started.
If the LHC will confirm the Tevatron exclusion,
the attention will move to the low mass region, where
the Higgs search is more difficult.

One of the most important production mechanisms of a light Higgs boson at hadron colliders is the
so called Higgs-strahlung process, i.e. the associated production of
the Higgs boson  together with a vector boson $V$ ($V=W^\pm$, $Z$).
At the Tevatron, this is the main search channel in the low mass region,
$m_H\ltap 140$ GeV, since the lepton(s) from the decay of the vector boson
provide the necessary background rejection.
At the LHC, this production channel has been considered less promising, due
to the large backgrounds. 
Recent studies \cite{Butterworth:2008iy}
have indicated that at large transverse momenta,
employing modern jet reconstruction and decomposition techniques,
\WH\ and \ZH\ production can be recovered as promising search modes
for a light Higgs boson at the LHC.

In order to fully exploit this channel it is important
to have accurate theoretical predictions for the production cross section
and the associated distributions.
Theoretical predictions with high precision demand in turn
detailed computations of radiative corrections.
The next-to-leading order (NLO) QCD corrections to \VH\ production
are the same as those
of the Drell-Yan process \cite{whnlo}.
Beyond NLO, the QCD corrections differ from those to the Drell-Yan
process \cite{Hamberg:1990np} by
contributions where the Higgs boson couples to the gluons through
a heavy-quark loop\footnote{In the case of $ZH$ production there are also gluon-gluon induced terms where both the $Z$ and $H$ bosons couple to a heavy-quark loop \cite{Brein:2003wg}.}.
The impact of these additional terms is, however, expected to be rather small.
Using the classical result of Ref.~\cite{Hamberg:1990np},
the computation of the NNLO inclusive cross section has been completed in
Ref.~\cite{Brein:2003wg}.
Also NLO electro-weak corrections have been evaluated \cite{Ciccolini:2003jy}.

The effect of NNLO QCD radiative corrections on the inclusive cross section is
relatively modest \cite{Brein:2003wg}. However, it is important to study how QCD corrections impact the accepted cross section and the relevant kinematical distributions.
This is particularly true when severe selection
cuts are applied, as it typically happens in Higgs boson searches.

The evaluation of higher-order QCD radiative corrections 
to hard-scattering processes 
is well known to be a hard task.
The presence of infrared singularities at intermediate stages 
of the calculation does not allow
a straightforward implementation of numerical techniques.
In particular, {\em fully differential} calculations at the 
NNLO involve a substantial amount of conceptual and technical
complications \cite{Kosower}-\cite{Czakon:2010td}.
In $e^+e^-$ collisions, NNLO differential cross sections are 
known only for 
two~\cite{Anastasiou:2004qd,Weinzierl:2006ij} and three jet 
production~\cite{threejets,Weinzierl:2008iv}.
In hadron--hadron collisions, fully differential cross sections have 
been computed only 
in the cases of Higgs production by gluon 
fusion \cite{Anastasiou:2004xq,Catani:2007vq,Grazzini:2008tf}
and of the Drell-Yan process \cite{Melnikov:2006di,Catani:2009sm,Catani:2010en}.

In this Letter we focus on \WH\ production and we present the fully exclusive NNLO QCD computation for this process. The calculation is performed by using the subtraction formalism of Ref.~\cite{Catani:2007vq}, and it is 
 based on an extension of the numerical program of Ref.~\cite{Catani:2009sm}.
We include finite-width effects, the leptonic decay of the $W$ boson, with its spin correlations, and the decay of the Higgs boson into a $b{\bar b}$ pair.
Only diagrams in which the Higgs boson is radiated by a $W$ boson are considered, i.e. we neglect
the contributions in which the Higgs boson couples to a heavy-quark loop.
Comparing our results to those of Ref.~\cite{Hirschi:2011pa}, where NLO predictions for $W\!H+{\rm jet}$, including these additional diagrams, are presented, we expect the neglected contributions to be
at the $1\%$ level or smaller.
When no cuts are applied, and the $W$ and $H$ are produced on shell, our numerical results agree with those obtained with the program {\tt VH@NNLO} \cite{Brein:2003wg}.

In the following we present an illustrative selection of numerical results for \WH\ production at the Tevatron and the LHC. We consider $u, d, s, c, b$ quarks in the initial state and we use the (unitarity constrained) CKM matrix elements $V_{ud} = 0.97428$, $V_{us} = 0.2253$, $V_{ub} = 0.00347$, $V_{cd} = 0.2252$, $V_{cs} = 0.97345$, $V_{cb} = 0.0410$ from the PDG 2010 \cite{Nakamura:2010zzi}.
As for the electroweak couplings, we use the so called $G_\mu$ scheme,
where the input parameters are $G_F$, $m_Z$, $m_W$. In particular we 
use the values
$G_F = 1.16637\times 10^{-5}$~GeV$^{-2}$,
$m_Z = 91.1876$~GeV, $m_W = 80.399$~GeV
and $\Gamma_W=2.085$~GeV.
Throughout the paper, we consider a SM Higgs boson with mass $m_H=120$ GeV and 
width $\Gamma_H=3.47$ MeV \cite{Dittmaier:2011ti}.
We compute the $H\to b{\bar b}$ decay at tree level
in the massless approximation,
and we normalize the $Hb{\bar b}$ Yukawa coupling such that $BR(H\to b{\bar b})=0.649$ \cite{Dittmaier:2011ti}.
We use the MSTW2008 \cite{Martin:2009iq} sets
of parton distributions, with
densities and $\as$ evaluated at each corresponding order
(i.e., we use $(n+1)$-loop $\as$ at N$^n$LO, with $n=0,1,2$).
The central values of the
renormalization and factorization scales are fixed to the value 
$\mu_R=\mu_F=m_W+m_H$.

We start the presentation of our results by considering \WH\ production at the Tevatron ($p{\bar p}$ collisions at $\sqrt{s}=1.96$ TeV).
We use the following cuts (see e.g. Ref.~\cite{Aaltonen:2009dh}).
We require the charged lepton to have transverse momentum $p_T^l > 20$ GeV and pseudorapidity $|\eta_l|< 2$, and the missing transverse momentum
of the event to fulfil $p_T^{\rm miss}> 20$ GeV.
Jets are reconstructed with the $k_T$ algorithm with $R=0.4$ \cite{ktalg}. We require exactly two jets with $p_T > 20$ GeV
and $|\eta|<2$, and at least one of them has to be a $b$ jet, with $|\eta|<1$.

\begin{table}[htbp]
\begin{center}
\begin{tabular}{|c|c|c|c|}
\hline
$\sigma$ (fb)& LO & NLO & NNLO\\
\hline
\hline
$\mu_F=\mu_R=(m_W+m_H)/2$ & $4.266 \pm 0.003$ & $4.840\pm 0.005$ & $4.788 \pm 0.013$\\
\hline
$\mu_F=\mu_R=m_W+m_H$ & $3.930\pm 0.003$ & $4.808\pm 0.004$ & $4.871 \pm 0.013$ \\
\hline
$\mu_F=\mu_R=2(m_W+m_H)$ & $3.639\pm 0.002$ & $4.738\pm 0.004$ & $4.908\pm 0.010$\\
\hline
\end{tabular}
\end{center}
\caption{{\em Cross sections for $p{\bar p}\to WH+X\to l\nu_lb{\bar b}+X$ at the Tevatron. The applied cuts are described in the text.}}
\label{tab:tev}
\end{table}

In Table~\ref{tab:tev} we report the accepted cross section\footnote{Throughout the paper, the errors
on the values of the cross sections and the error bars in the plots refer to an
estimate of the numerical errors in the Monte Carlo integration.}
at LO, NLO and NNLO in the case of
three different scale choices around the central value $\mu_F=\mu_R=m_W+m_H$.
The impact of NLO (NNLO) corrections ranges from $+13\%$ to $+30\%$ ($-1\%$ to $+4\%$), depending on the scale choice.
The scale dependence is at the level of about $\pm 1\%$ both at NLO and NNLO.

\begin{figure}[htb]
\begin{center}
\begin{tabular}{c}
\epsfxsize=10truecm
\epsffile{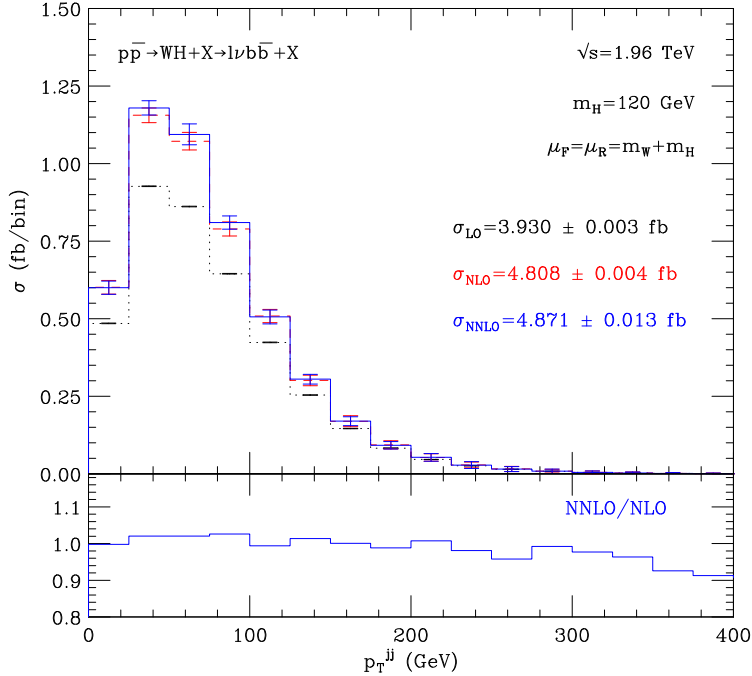}\\
\end{tabular}
\end{center}
\caption{\label{fig:ptjj}
{\em Transverse-momentum spectrum of the dijet system for
$p{\bar p}\to WH+X\to l\nu b{\bar b}+X$ at the Tevatron
at LO (dots), NLO (dashes) and NNLO (solid). The applied cuts are described in the text.}}
\end{figure}

In Fig.~\ref{fig:ptjj} we show the transverse-momentum spectrum of the dijet system at LO, NLO, NNLO.
The lower panel of the figure shows the ratio NNLO/NLO.
We see that the shape of the spectrum is rather stable, when going from NLO to NNLO, within the statistical uncertainties.

We now consider \WH\ production at the LHC ($\sqrt{s}=14$ TeV).
We follow the selection strategy of Ref.~\cite{Butterworth:2008iy} (see also \cite{atlasnote}):
the Higgs boson is searched for at large transverse momenta through its decay into a collimated $b{\bar b}$ pair.
We require the charged lepton to have $p_T^l > 30$ GeV and $|\eta_l|< 2.5$, and the missing transverse momentum
of the event to fulfil $p_T^{\rm miss}> 30$ GeV.
We also require the $W$ boson to have $p_T^W>200$ GeV.
Jets are reconstructed with the Cambridge/Aachen algorithm \cite{caalg}, with $R=1.2$.
One of the jets (fat jet) must have $p_T^J>200$ GeV and $|\eta_J|<2.5$ and must contain the $b{\bar b}$ pair.
There should not be other jets with $p_T>20$ GeV and $|\eta|< 5$.

\begin{table}[htbp]
\begin{center}
\begin{tabular}{|c|c|c|c|}
\hline
$\sigma$ (fb)& LO & NLO & NNLO\\
\hline
\hline
$\mu_F=\mu_R=(m_W+m_H)/2$ & $2.640\pm 0.002$ & $1.275\pm 0.003$ & $1.193 \pm 0.017$\\
\hline
$\mu_F=\mu_R=m_W+m_H$ & $2.617\pm 0.003$ & $1.487\pm 0.003$ & $1.263 \pm 0.014$ \\
\hline
$\mu_F=\mu_R=2(m_W+m_H)$ & $2.584\pm 0.003$ & $1.663\pm 0.002$ & $1.346\pm 0.013$\\
\hline
\end{tabular}
\end{center}
\caption{{\em Cross sections for $pp\to WH+X\to l\nu_lb{\bar b}+X$ at the LHC. The applied cuts are described in the text.}}
\label{tab:atlas}
\end{table}

In Table \ref{tab:atlas} we report the corresponding accepted cross sections at LO, NLO and NNLO.
We see that the impact of NLO and NNLO corrections is negative and larger than at the Tevatron.
The NLO (NNLO) effect ranges from $-52\%$ to $-36\%$ ($-6\%$ to $-19\%$), depending on the scale choice.
The scale dependence goes from about $\pm 13\%$ at NLO to about $\pm 6\%$ at NNLO.
The NLO and NNLO accepted cross sections are compatible within the scale uncertainties.

\begin{figure}[htb]
\begin{center}
\begin{tabular}{c}
\epsfxsize=10truecm
\epsffile{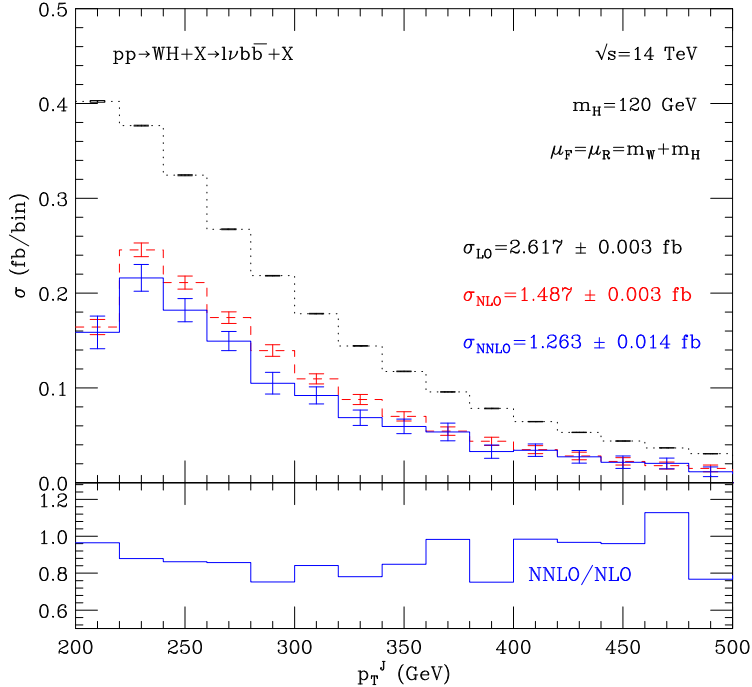}\\
\end{tabular}
\end{center}
\caption{\label{fig:ptfat}
{\em Transverse-momentum spectra of the fat jet for
$pp\to WH+X\to l\nu b{\bar b}+X$ at the LHC
at LO (dots), NLO (dashes) and NNLO (solid). The applied cuts are described in the text.}}
\end{figure}

In Fig.~\ref{fig:ptfat} we show the transverse-momentum spectrum of the fat jet at LO, NLO and NNLO.
The lower panel shows the ratio NNLO/NLO. We see that the shape of the distribution is relatively stable when going from NLO to NNLO.

We add few comments on the different impact of QCD radiative corrections at the Tevatron and at the LHC \cite{Catani:2001cr}.
At the Tevatron, the invariant mass of the \WH\ system is $M_{WH}\sim m_W+m_H$. The typical scale of the accompanying QCD radiation is of the order of about $\langle 1-z \rangle\,M_{WH}$ where $\langle 1-z\rangle=\langle 1-M_{WH}^2/{\hat s}\rangle$ is the average distance
from the partonic threshold. The effect of the veto on additional jets is thus marginal if the jet veto scale, $p_T^{\rm veto}$, is of the order of $\langle 1-z \rangle\,M_{WH}$. In this case the perturbative expansion
appears under good control. The situation at the LHC is different in two respects. First, the invariant mass of the \WH\ system is larger, due to the high $p_T$ required for the $W$ and the Higgs candidate.
Second, the typical distance from the partonic threshold is larger, i.e. $\langle 1-z \rangle$ is larger than at the Tevatron, due to the increased $\sqrt{s}$. As a consequence, a stringent veto on the radiation recoiling against the \WH\ system
spoils the cancellation of the infrared singularities in real and virtual corrections, and contributions enhanced
by the logarithm of the ratio $(1-z)M_{WH}/p_T^{\rm veto}$ are definitely relevant.
We have checked that the reduction in the accepted cross section is in fact due to the jet veto, the impact of QCD corrections being positive if the jet veto is removed.

We have illustrated a calculation of the NNLO cross section for \WH\ production in hadron collisions. The calculation is implemented in a parton level event generator and allows us to apply arbitrary kinematical cuts on the $W$ and $H$ decay products
as well as on the accompanying QCD radiation. We have studied the impact of NNLO QCD corrections in two typical cases
at the Tevatron and the LHC. At the Tevatron, the perturbative expansion appears under good control. At the LHC, by searching
for events where the Higgs boson is boosted at high $p_T$, the impact of QCD corrections is more sizeable, and the stability
of the fixed-order calculation is challenged.
More detailed studies, along the lines of Refs.~\cite{Anastasiou:2008ik,Anastasiou:2009bt}, are needed in order to assess the relevance of these fixed-order perturbative result.

\noindent {\bf Acknowledgements.}
We are grateful to Rikkert Frederix for allowing us to cross check our NLO results for \WH\ + jet with those of the numerical program of Ref.~\cite{Hirschi:2011pa}.
We thank Robert Harlander for sending us a version of {\tt VH@NNLO}, Stefano Catani, Nicolas Greiner and Stefano Pozzorini for helpful
discussions and comments on the manuscript. We also thank Craig Group, Tom Junk and Giacinto Piacquadio for discussions on the experimental analysis at the Tevatron and the LHC.
This work has been supported in part by the European Commission through the 'LHCPhenoNet' Initial Training Network PITN-GA-2010-264564.

\end{document}